\documentclass[fleqn,usenatbib,dvipdfmx]{mnras}

\usepackage[T1]{fontenc}
\usepackage{ae,aecompl}

\usepackage{graphicx}	
\usepackage{amsmath}	
\usepackage{amssymb}	
\usepackage{ulem}

\title[
Energy generation rates and PISN
]{
Exact and approximate expressions of energy generation rates
and their impact on the explosion properties of Pair Instability Supernovae
}

\author[K. Takahashi et al.]{
Koh Takahashi$^1$\thanks{E-mail: ktakahashi@astron.s.u-tokyo.ac.jp (KT)},
Takashi Yoshida$^{1, 2}$,
Hideyuki Umeda$^1$,
Kohsuke Sumiyoshi$^3$,
\newauthor
and
Shoichi Yamada$^{4, 5}$
\\
$^1$ Department of Astronomy, Graduate School of Science, University of Tokyo, Tokyo 113-0033, Japan\\
$^2$ Yukawa Institute for Theoretical Physics, Kyoto University, Oiwake-cho, Kitashirakawa, Sakyo-ku, Kyoto, 606-8502, Japan\\
$^3$ Physics Group, Numazu College of Technology, Ooka 3600, Numazu, Shizuoka 410-8501\\
$^4$ School of Advanced Science and Engineering, Waseda University, 3-4-1, Okubo, Shinjuku, Tokyo 169-8555, Japan\\
$^5$ Advanced Research Institute for Science and Engineering, Waseda University, 3-4-1, Okubo, Shinjuku, Tokyo 169-8555, Japan\\
}

\date{Accepted XXX. Received YYY}

\pubyear{2015}

\begin{document}
\label{firstpage}
\pagerange{\pageref{firstpage}--\pageref{lastpage}}
\maketitle

\begin{abstract}
Energetics of nuclear reaction is fundamentally important to understand the mechanism of pair instability supernovae (PISNe).
Based on the hydrodynamic equations and thermodynamic relations, we derive exact expressions for energy conservation suitable to be solved in simulation.
We also show that some formulae commonly used in the literature are obtained as approximations of the exact expressions.
We simulate the evolution of very massive stars of $\sim$100-320 M$_\odot$ with zero- and 1/10 Z$_\odot$,
and calculate further explosions as PISNe, applying each of the exact and approximate formulae.
The calculations demonstrate that the explosion properties of PISN, such as the mass range, the $^{56}$Ni yield, and the explosion energy,
are significantly affected by applying the different energy generation rates.
We discuss how these results affect the estimate of the PISN detection rate, which depends on the theoretical predictions of such explosion properties.
\end{abstract}




\section{Introduction}

Energy generation rates, which are commonly denoted by $\epsilon$ in the equation of energy conservation,
are essentially important terms that control the time evolution of a star throughout its life.
Based on the equation of one dimensional hydrodynamics, classical textbooks formalize energy conservation with three kinds of energy generation rates:
\begin{eqnarray}
	\frac{ {\rm d} L }{ {\rm d} M }	= \epsilon_{grv}
							+ \epsilon_{nuc}
							 - \epsilon_{\nu}, \label{stellar_eng}
\end{eqnarray}
where $L$ and $M \equiv \int^{r}_{0} 4\pi r^2 \rho \mathrm{d}r$ are the luminosity and the enclosed mass at the radius $r$,
$\epsilon_{grv}$, $\epsilon_{nuc}$, and $\epsilon_{\nu}$ are 
the so-called gravothermal energy generation rate,
the nuclear energy generation rate,
and the neutrino cooling rate, respectively \citep{chiu68, cox68, kippenhahn&weigert90, iben13}.

In order to obtain the complete expression of this equation, definitions of individual energy generation rates should be given explicitly.
In the literature, however, such definitions are often omitted or given with some implicit assumptions.
For example, as noted in \citet{cox68}, if one firstly defines $\epsilon_{grv}$ as
\begin{eqnarray*}
	\epsilon_{grv} \equiv - \frac{ \mathrm{D}e }{ \mathrm{D}t }
	 - p \frac{ \mathrm{D}(1/\rho) }{ \mathrm{D}t }
\end{eqnarray*}
using the Lagrangian time derivative $\mathrm{D}/\mathrm{D}{t}$,
the specific internal energy density $e$, the pressure $p$, and the density $\rho$ of the gas,
then $\epsilon_{grv}$ is equivalently given by
\begin{eqnarray*}
	\epsilon_{grv} = - T \frac{ \mathrm{D}s }{ \mathrm{D}t }  - \frac{1}{m_u} \sum_{particles} \mu_i \frac{ \mathrm{D}Y_i }{ \mathrm{D}t },
\end{eqnarray*}
where $T$, $s$, and $m_u$ are the temperature, the specific entropy, and the atomic mass unit,
and $\mu_i$ and $Y_i$ are the chemical potential and the mole fraction of $i$-th particle, respectively.
If one assumes that the second term on the right hand side is negligibly small, this equation becomes
\begin{eqnarray*}
	\epsilon_{grv} = - T \frac{ \mathrm{D}s }{ \mathrm{D}t },
\end{eqnarray*}
which is the formula given in \citet{chiu68} and in \citet{kippenhahn&weigert90}
(equivalent formalisms are used in, e.g., \citet{hayashi+62, paxton+11}, Chieffi 2015, private communication).

As for the nuclear energy generation rate, the general definition is more difficult to find\footnote{Eq.~8.5 in \citet{kippenhahn&weigert90} may provide this.}.
One may define $\epsilon_{nuc}$ using the Q-value and the reaction rate $\lambda$
\begin{eqnarray*}
	\epsilon_{nuc} \equiv \sum_{reactions} Q_i \lambda_i
\end{eqnarray*}
as a straightforward generalization of that for the $pp$ chain or the $CNO$ cycle that are often discussed in detail.
In this case, the Q-value must contain the rest masses (and the average energy for neutrino) of {\it all} particles involved in the reaction,
and the summation must run for {\it all} reactions occurring.
If little attention is paid to these points, physically incorrect energy generation rates might produce erroneous results in simulation.

As a demonstration, we show the results of explosion simulation of pair instability supernova (PISN).
Two key reactions are responsible for the explosion \citep{barkat+67, rakavy+67}.
The first is a creation reaction of electron-positron pair that reduces the pressure and induces the dynamical collapse of a CO core,
and the second is a thermonuclear reaction of oxygen burning that effectively heats the core and supplies energy to reverse the motion from collapse to explosion.
We discuss how these energetics can be treated in simulation and
what is missing in approximate formulae which seems to have been employed in the stellar evolution community.

Theoretical investigation of PISN has been driven by an expectation that 
their outstanding properties, such as a high explosion energy and a peculiar yield pattern, may enable the
observational identification of the hidden properties of the far-away universe \citep[e.g.,][]{umeda&nomoto02, heger&woosley02}.
In particular, owing to the high explosion energy and/or large $^{56}$Ni yield estimated which make the explosion luminous,
PISN is one of the most promising candidates as an observable high redshift object \citep{scannapieco+05, kasen+11, kozyreva+14a, kozyreva+14b, chatzopoulos+15}.
Recently, the possibility of direct detection of high-redshift PISNe by next generation telescopes has been actively debated \citep{whalen+13, smidt+15}.
In the estimate of the detection rate, accurate determination of the intrinsic properties of PISN is of prime importance.
Therefore, we also discuss how errors caused by incorrect energy generation rates affect the rate estimate of PISN.

In the next section, we derive four exact expressions of the energy generation rates based on the hydrodynamic equations and thermodynamic relations (\S 2.1).
In addition to the exact formulae, two approximate expressions which seem to have been applied in stellar simulations, are obtained (\S 2.2).
Settings of simulations are summarized in \S 3.
We calculate the evolution and subsequent explosion of very massive stars of $\sim$100-320 M$_\odot$, which have a metallicity of zero- or 1/10 Z$_\odot$.
For the demonstration, we mainly apply two different expressions of the energy generation rates, one exact and the other approximate, to the explosion calculation.
Some features are common among these sets of calculations (\S 4.1),
while application of the approximate formula causes significant errors in PISN properties (\S 4.2).
How these errors affect the rate estimate of PISN is discussed in \S 5.
Conclusions are given in \S 6.

\section{Energy generation rates in the stellar equation}

\subsection{Exact expressions}

We begin the discussion with the energy conservation for perfect fluid,
\begin{eqnarray}
	\frac{ \mathrm{D}e^{rel} }{ \mathrm{D}t } =
								- p\frac{ \mathrm{D}(1/\rho_b) }{ \mathrm{D}t }
								- \frac{ 1 }{ \rho_b } {\rm div} \mbox{\boldmath $\theta$}
								- \epsilon_\nu,
\end{eqnarray}
where
$\rho_b \equiv m_u n_b$ is the baryon density defined as the product of the atomic mass unit and the baryon number density,
$\rho_b e^{rel}$ is the relativistic internal energy per unit volume, $p$ is the pressure, and $\mbox{\boldmath $\theta$}$ is the energy flux.
The last term on the right hand side contains contributions from all processes that emit neutrinos,
such as bremsstrahlung, plasmon decay, electron capture, beta decay, etc.
It is worth noting that this energy equation does not include the term of $\epsilon_{nuc}$.
This is because nuclear reactions just transform the form of the internal energy from the rest mass to thermal motion of gas particles.
On the other hand, $\epsilon_{\nu}$ is included since neutrinos are not in thermal equilibrium with stellar gas
and the energy of neutrinos are excluded from the internal energy.

Firstly, we impose spherical symmetry and take the enclosed baryon mass $M_b \equiv \int^{r}_{0} 4\pi r^2 \rho_b \mathrm{d}r$ as the coordinate of the system
to obtain a suitable expression for the stellar evolution calculation, which is not essential point, though.
Then, one may obtain
\begin{eqnarray}
	\frac{ \mathrm{d}L }{ \mathrm{d}M_b } = 
								- \frac{ \mathrm{D}e^{rel} }{ \mathrm{D}t } 
								- p\frac{ \mathrm{D}(1/\rho_b) }{ \mathrm{D}t }
								- \epsilon_\nu, \label{rep_base}
\end{eqnarray}
where the relation $L = 4\pi r^2 \theta_r$ is used.
This equation coincides with eq.(\ref{stellar_eng}), if $\epsilon_{grv}$ and $\epsilon_{nuc}$ are correctly defined as
\begin{eqnarray}
	\epsilon_{grv}^{base}	&\equiv&	- \frac{ {\rm D}e^{rel} }{ {\rm D}t } 
	 - p\frac{ {\rm D}(1/\rho_b) }{ {\rm D}t }, \\
	\epsilon_{nuc}^{base}	&\equiv& 0.
\end{eqnarray}
Hereafter we refer to these definitions as the {\it base expression} of energy conservation.

Next, we define the thermal component of the internal energy density as
\begin{eqnarray}
	\rho_b e^{therm} \equiv \rho_b e^{rel} - \rho c^2, \label{thermeng}
\end{eqnarray}
in which $\rho$ is the rest mass density (\ref{rest_mass}).
According to eq.(\ref{rest_mass2}), the change of the rest mass per baryon is expressed as
\begin{eqnarray}
	{\rm d} \Bigl( \frac{ \rho c^2}{\rho_b} \Bigl) = \frac{1}{m_u} \Bigl[
				\sum_{ion} m_i c^2 {\rm d}Y_i
				 + m_e c^2 {\rm d}Y_e
				  + 2m_e c^2 {\rm d}Y_{e^+} \Bigl], \label{rest_mass3}
\end{eqnarray}
where
$m_i$ and $Y_i$ are the rest mass and the mole fraction of $i$-th ion,
$m_e$ is the electron mass, $Y_e = Y_{e^-} - Y_{e^+}$ is the net electron mole fraction,
and $Y_{e^-}$ and $Y_{e^+}$ are the electron and positron mole fractions (see the appendix~\ref{app1} for detail definitions).
Equating eqs.(\ref{rep_base},\ref{thermeng},\ref{rest_mass3}), one obtains an alternative expression of energy conservation as
\begin{eqnarray}
	\frac{ \mathrm{d}L }{ \mathrm{d}M_b } =  \epsilon^{reac}_{grv}
								+ \epsilon^{reac}_{nuc}
								- \epsilon_\nu,
\end{eqnarray}
where we momentarily identify the energy generation rates as
\begin{eqnarray}
	\epsilon_{grv}^{reac}  &\equiv& - \frac{ {\rm D}e^{therm} }{ {\rm D}t } - p\frac{ {\rm D}(1/\rho_b) }{ {\rm D}t } \label{rep_reac_grv}\\
	\epsilon_{nuc}^{reac} &\equiv& - \frac{1}{m_u}
					\Bigl[		\sum_{ion} m_i c^2 \frac{ {\rm D}Y_i }{ {\rm D}t }
							+ m_e c^2 \frac{ {\rm D} Y_e}{ {\rm D}t }
							+ 2 m_e c^2 \frac{ {\rm D} Y_{e^+} }{ {\rm D}t } \Bigl]. \label{rep_reac_nuc}
\end{eqnarray}
This expression, though it is equivalent to the base expression, is more suitable for intuitive understanding of the effect of reactions,
since the thermal part of the internal energy is affected by $\epsilon^{reac}_{nuc}$, which is directly related to the compositional change due to reactions.
On the other hand, since one has to evaluate Fermi-Dirac integrals to obtain $Y_{e^+}$, this formula is not feasible to be applied to a numerical simulation.

One practical solution to take account of the change of the rest mass of pair electrons and positrons
is to include the rest mass of that particular particles into the internal energy of the EOS \citep[e.g.,][]{blinnikov+96, timmes&swesty00}.
Thus, the internal energy in a stellar code is often defined as
\begin{eqnarray}
	\rho_b e^{therm+pair} &\equiv&	\rho_b e^{therm} + \rho_{pair}c^2 \\
					&=&		\rho_b e^{therm} + 2 m_e c^2 n_{e^+},
\end{eqnarray}
then the equivalently exact expression, which we call the {\it reaction expression}, is derived:
\begin{eqnarray}
	\epsilon_{grv}^{reac}  &\equiv& - \frac{ {\rm D}e^{therm+pair} }{ {\rm D}t } - p\frac{ {\rm D}(1/\rho_b) }{ {\rm D}t } \label{rep_prac_grv} \\
	\epsilon_{nuc}^{reac} &\equiv& - \frac{1}{m_u}
					\Bigl[		\sum_{ion} m_i c^2 \frac{ {\rm D}Y_i }{ {\rm D}t }
							+ m_e c^2 \frac{ {\rm D} Y_e}{ {\rm D}t } \Bigl]. \label{rep_prac_nuc}
\end{eqnarray}
This expression does no longer include the term of $Y_{e^+}$.
Similar treatments can be found in some stellar codes that is used for massive stellar evolution calculation
(private communication, Woosley 2015, Heger 2015, Langer 2015, Timmes 2015).

The other way to eliminate the term of $Y_{e^+}$ is to treat the entropy equation instead.
This approach guarantees a fluid element evolves adiabatically when the energy flux and reactions are negligible.
Moreover, this has an advantage to calculate a thermal structure of degenerate objects like white dwarfs,
in which pressure and internal energy are scarcely dependent on temperature.

The entropy equation is obtained by equating the one-dimensional energy conservation (eq.\ref{rep_base}) with the second law of thermodynamics in the form of specific density,
\begin{eqnarray}
	{\rm d} e^{rel} = T{\rm d}s - p{\rm d} \Bigl( \frac{1}{\rho_b} \Bigl)
					+ \frac{1}{m_u} \sum_{particles} \mu^{rel}_i {\rm d}Y_i,
\end{eqnarray}
where $T$ is the temperature, $\rho_b s$ is the entropy per unit volume,
and $\mu^{rel}_i$ is the relativistic chemical potential (including the rest mass) of $i$-th particle (see appendix~\ref{app2}).
Since the reaction equilibrium is always established for the pair reactions of $\gamma + \gamma \rightleftharpoons e^- + e^+$,
the relation $\mu^{rel}_{e^+} = - \mu^{rel}_{e^-}$ is satisfied.
Then one obtains
\begin{eqnarray}
	\frac{ {\rm d}L }{ {\rm d}M_b } 
	=	- T \frac{ {\rm D}s }{ {\rm D}t }
		- \frac{1}{m_u}	\Bigl[	\sum_{ion} \mu^{rel}_i \frac{ {\rm D}Y_i }{ {\rm D}t }
					+\mu^{rel}_e \frac{ {\rm D} Y_e}{ {\rm D}t } \Bigl]
		- \epsilon_\nu.
\end{eqnarray}
This equation does not include the term of $Y_{e^+}$, and thus can be evaluated only using ionic mole fractions.
As for the definitions of the energy generation rates, an ambiguity exists.
If one defines $\epsilon_{nuc}^{ent}$ to account for the heat generated by reactions, the following {\it entropy expression} can be defined,
\begin{eqnarray}
	\epsilon_{grv}^{ent}	&\equiv&- T \frac{ {\rm D}s }{ {\rm D}t }, \label{rep_ent_grv}\\
	\epsilon_{nuc}^{ent}	&\equiv&- \frac{1}{m_u}	\Bigl[		\sum_{ion} \mu^{rel}_i \frac{ {\rm D}Y_i }{ {\rm D}t }
											+\mu^{rel}_e \frac{ {\rm D} Y_e}{ {\rm D}t } \Bigl]. \label{rep_ent_nuc}
\end{eqnarray}

\subsection{Approximate expressions}

In addition to the above equivalently exact expressions, here we derive two approximate expressions,
which are often found in the literature and, we are afraid, may have been employed in actual simulations.

Firstly, if one neglects the last term of eq.(\ref{rep_reac_nuc}), then the reaction expression becomes
\begin{eqnarray}
	\tilde \epsilon_{grv}^{reac}		&\equiv& - \frac{ {\rm D}e^{therm} }{ {\rm D}t }
									- p\frac{ {\rm D}(1/\rho_b) }{ {\rm D}t }, \label{rep_reac_grv2} \\
	\tilde \epsilon_{nuc}^{reac}	&\equiv& - \frac{1}{m_u}	\Bigl[		\sum_{ion} m_i c^2 \frac{ {\rm D}Y_i }{ {\rm D}t }
									+ m_e c^2 \frac{ {\rm D} Y_e}{ {\rm D}t } \Bigl]. \label{rep_reac_nuc2}
\end{eqnarray}
This {\it approximate reaction expression} is easy to evaluate similar to $\epsilon^{ent}_{nuc}$ and
coincides with the exact one if positron is essentially non-existent.
Thus, to investigate evolution of low mass stars or early stages of massive stellar evolution, this expression provides accurate enough energy generation rates,
and this is why the expression is utilized in some stellar evolution codes (e.g. in GENEC, Meynet 2015, private communication).
On the other hand, this expression overestimates the nuclear energy generation rate when electron-positron pairs are created, 
since the neglected positron term accounts for the energy reduction due to pair creation.
Correspondingly, the nuclear energy generation rate is underestimated when the pair annihilation occurs.

Next, the other approximate expression is derived from the exact entropy expression,
in which the thermal contribution of the chemical potential $\mu_i^{therm} \equiv \mu^{rel}_i - m_i c^2$ is neglected:
\begin{eqnarray}
	\tilde \epsilon_{grv}^{ent}	&\equiv&- T \frac{ {\rm D}s }{ {\rm D}t }, \label{rep_ent_grv2}\\
	\tilde \epsilon_{nuc}^{ent}	&\equiv& - \frac{1}{m_u}	\Bigl[		\sum_{ion} m_i c^2 \frac{ {\rm D}Y_i }{ {\rm D}t }
													+ m_e c^2 \frac{ {\rm D} Y_e}{ {\rm D}t } \Bigl]. \label{rep_ent_nuc2}
\end{eqnarray}
This {\it approximate entropy expression}, hence, coincides with the exact one
when the thermal contribution to the chemical potential is negligibly small compared with the rest mass.
Meanwhile, our calculations show that this expression overestimates the nuclear energy generation rate in general.

It is noteworthy that, though totally different assumptions are treated in each expression,
the appearance of the nuclear energy generation rates of
the exact reaction expression (eq.\ref{rep_prac_nuc}),
the approximate reaction expression (eq.\ref{rep_reac_nuc2}), and
the approximate entropy expression (eq.\ref{rep_ent_nuc2}) are in complete agreement with each other.
Thus, it is necessary to exhibit the definition of both the so-called gravothermal energy generation rate and the nuclear energy generation rate
in order to illustrate what kind of physics are really treated in the simulation.

\section{COMPUTATIONAL SETTINGS}

We calculate the evolution and subsequent explosion of non-rotating very massive stars having initial masses of $\sim$100-320 $M_\odot$ for two metallicities of zero and 1/10 $Z_\odot$.
With a stellar evolution code \citep{takahashi+14}, hydrostatic evolution is calculated from the hydrogen burning stage until the end of the helium burning stage.
Later evolution including explosion is calculated by a time-implicit general-relativistic Lagrangian hydrodynamic code described in \citet{yamada97}.

\begin{table}
\centering
	\caption{Isotopes included in the stellar evolution code (left) and in the hydrodynamic code (right).}
	\label{tab_isotope}
	\begin{tabular}{ccc|ccc}
	\hline
	Element & \multicolumn{2}{c}{mass number} & Element & \multicolumn{2}{c}{mass number} \\
	\hline
	n	&	1		&	1		&	Ar	&	33-42	&	34-40	\\
	H	&	1-3		&	1-3		&	K	&	36-43	&	37-41	\\
	He	&	3-4		&	3-4		&	Ca	&	37-48	&	38-43	\\
	Li	&	6-7		&	6-7		&	Sc	&	40-49	&	41-45	\\
	Be	&	7-9		&	7-9		&	Ti	&	41-51	&	43-48	\\
	B	&	8-11		&	8-11		&	V	&	44-52	&	45-51	\\
	C	&	11-14	&	12-13	&	Cr	&	46-55	&	47-54	\\
	N	&	12-15	&	13-15	&	Mn	&	48-56	&	49-55	\\
	O	&	13-20	&	14-18	&	Fe	&	50-61	&	51-58	\\
	F	&	17-21	&	17-19	&	Co	&	54-62	&	53-59	\\
	Ne	&	18-24	&	18-22	&	Ni	&	56-66	&	55-62	\\
	Na	&	20-25	&	21-23	&	Cu	&	59-67	&	57-63	\\
	Mg	&	21-27	&	22-26	&	Zn	&	62-70	&	60-64	\\
	Al	&	23-29	&	25-27	&	Ga	&	65-73	&	-	\\
	Si	&	24-32	&	26-32	&	Ge	&	69-76	&	-	\\
	P	&	27-34	&	29-33	&	As	&	71-77	&	-	\\
	S	&	29-36	&	30-36	&	Se	&	73-79	&	-	\\
	Cl	&	31-38	&	33-37	&	Br	&	76-80	&	-	\\
	\hline
	\end{tabular}
\end{table}

In order to get a smooth transition from the evolution code to the hydrodynamic code, the same EOS is implemented.
In the EOS, four species of particles, photon, ions, electron, and positron are considered.
Photon is expressed as a black body radiation.
For ions, sum of mole fractions $Y_I \equiv \sum_{ions} Y_i$ is used to calculate the pressure and the entropy,
while full composition is used for chemical potential calculation.
For the electron-positron part, reaction equilibrium of $\gamma + \gamma \rightleftharpoons e^- + e^+$ is always assumed,
and analytic approximations for general Fermi-Dirac integrals are used \citep{blinnikov+96}.
Moreover, the same nuclear reaction network and neutrino cooling formulae \citep{itoh+96} are applied to the two codes.
The number of isotopes are 260 for the stellar evolution code and 153 for the hydrodynamic code (Table \ref{tab_isotope}).
The number for the hydrodynamic code is chosen to estimate the energy generation rate accurately.

In the stellar evolution code, mass loss is taken into account for 1/10 Z$_\odot$ stars.
The standard case description in \citet{yoshida&umeda11} is applied, thus,
the metallicity dependence of $(Z/Z_\odot)^{0.64}$ is used for the red giant mass loss ($T_{eff}$ < 12 000K).
The formula for the red giant mass loss is changed from the mass loss rate by \citet{deJager+88} to the rate by \citet{nieuwenhuijzen&deJager90},
to avoid a rapid rate increase for luminous stars that is introduced due to a numerical reason \citep[see discussions in][]{muijres+12}.
With this description, severe mass loss possibly reduces the CO core mass of a very massive star, dismissing the star from the PISN mass range.
Since mass loss rate is larger for higher metallicity, the metallicity of 1/10 Z$_\odot$ is close to a maximum metallicity for PISNe to occur \citep{langer+07,yoshida&umeda11, yusof+13}.

Meanwhile, mass loss is neglected for zero metallicity stars, as assumed in a previous work \citep{umeda&nomoto02}.
This is because, the surface of a zero metallicity star does not contain metals such as iron,
which account for photon absorption that accelerates the stellar wind of metal rich stars.
However, the mechanism, and thus the efficiency, of mass loss of very massive stars are highly uncertain.
Some other possibilities that activate even for zero metallicity stars have been discussed \citep[e.g., pulsation mass loss, ][]{barafe+01, sonoi&umeda12, moriya&langer15}.
Therefore, present sets of calculations with different metallicities would be regarded as two sets of calculations, in which efficient mass loss is taken into account or not.

For the PISN explosion simulation, mainly two different expressions of energy generation rates are applied.
The first one is the exact entropy expression (eqs.\ref{rep_ent_grv},\ref{rep_ent_nuc}).
Hereafter, we refer to this group of calculations as the case A calculations.
To be accurate, the hydrodynamic code solves two equations that are equivalent to the entropy expression,
\begin{eqnarray*}
	\frac{ \mathrm{D}e^{rel} }{ \mathrm{D}t } &=&
								- p\frac{ \mathrm{D}(1/\rho_b) }{ \mathrm{D}t }
								- \epsilon_\nu, \\
	{\rm d}e^{rel} &=& T{\rm d}s
					- p{\rm d} \Bigl( \frac{1}{\rho_b} \Bigl)
					+ \frac{1}{m_u}	\Bigl[		\sum_{ion} \mu^{rel}_i {\rm d}Y_i 
										+\mu^{rel}_e {\rm d} Y_e \Bigl].
\end{eqnarray*}
The second equation is used to determine the specific entropy that is used as the independent variable of the EOS.
Note that the luminosity term is neglected in the energy equation, since the other terms overwhelm the luminosity term for the short timescale evolution.
The second expression is the approximate entropy expression (eqs.\ref{rep_ent_grv2},\ref{rep_ent_nuc2}).
Similarly to the case A,
\begin{eqnarray*}
	\frac{ \mathrm{D}e^{rel} }{ \mathrm{D}t } &=&
								- p\frac{ \mathrm{D}(1/\rho_b) }{ \mathrm{D}t }
								- \epsilon_\nu \\
	{\rm d}e^{rel} &=& T{\rm d}s
					- p{\rm d} \Bigl( \frac{1}{\rho_b} \Bigl)
					+ \frac{1}{m_u}	\Bigl[		\sum_{ion} m_i c^2 {\rm d}Y_i 
										+ m_e c^2 {\rm d} Y_e \Bigl]
\end{eqnarray*}
are solved.
These calculations are referred to the case B calculations.

In order to determine the minimum mass of PISN, additional hydrodynamic calculations are conducted for less massive stars of $\sim$100-140 M$_\odot$.
The distributions obtained by an explosion calculation is mapped onto the time-explicit Lagrangian hydrodynamic code \citep{colella&woodward84, umeda&nomoto02},
and further expansion is calculated.
The same EOS, nuclear reaction network, and neutrino cooling formulae are implemented.
Based on the calculation, a star whose central region does not fall back into the center at $10^5$ sec after the explosion is considered to become a PISN.
Otherwise, the fate is identified as a pulsational pair instability supernova \citep[PPISN,][]{barkat+67, heger&woosley02}.

In addition to the calculation based on the entropy expressions,
we have calculated explosions applying the reaction expression (eqs.\ref{rep_prac_grv},\ref{rep_prac_nuc})
to make a comparison between the two exact expressions.
In this case, two equations,
\begin{eqnarray*}
	\frac{ \mathrm{D}e^{rel} }{ \mathrm{D}t } &=&
								- p\frac{ \mathrm{D}(1/\rho_b) }{ \mathrm{D}t }
								- \epsilon_\nu \\
	\mathrm{d}e^{rel} &=&			\mathrm{d}e^{therm+pair}
								+ \frac{1}{m_u}	\Bigl[		\sum_{ion} m_i c^2 {\rm d}Y_i
													+ m_e c^2 {\rm d} Y_e \Bigl],
\end{eqnarray*}
are solved, where the thermal part of the internal energy is calculated to determine the specific entropy by $e^{therm+pair} = e^{therm+pair}(s, \rho_b, Y_i, Y_e)$.
The result shows good agreement with the result obtained by adopting the exact entropy expression.
The comparison is briefly discussed in \S 4.3.

\section{Result}

Properties of calculated models are listed in Table~\ref{stars}.
Difference between the initial mass and the final mass shows how amount of mass is ejected by mass loss during the evolution.
Since mass loss for a zero metallicity star is neglected, the core is surrounded by a hydrogen envelope throughout the evolution.
On the other hand, strong mass loss during the red giant phase strips the whole envelope of a 1/10 Z$_\odot$ star, then,
only a bare helium star remains with this metallicity.
As shown later, existence of the envelope at the moment of the explosion affects the minimum mass for PISN.

The fate of a very massive star mainly depends on the core mass.
During the hydrogen and helium burning stages, a central region of a very massive star becomes convective,
and the star forms a CO core of about a half of its initial mass.
The He or the CO core mass is determined as the mass coordinates, at which the mass fractions of hydrogen or helium becomes less than 0.1.
\begin{table*}
\centering
	\caption{
	Properties of calculated models.
	$M_{ini}$, $M_{fin}$, $M_{He}$, and $M_{CO}$ are the initial mass, the final mass, the helium core mass, and the carbon-oxygen core mass,
	C/O is the ratio between mass fractions of carbon and oxygen in the CO core,
	$E_{tot}$, $M_{^{56}Ni}$, and $T_{max}$ are the explosion energy, the ejected $^{56}$Ni mass, and the maximum temperature reached during the explosion, respectively.
	The fate is specified from PPISN, PISN, or black hole formation (BH).
	All of the masses are normalized by solar mass, the total energy is in $10^{51}$ erg, and $T_{max}$ is in kelvin.
	}
	\label{stars}
	\begin{tabular}{ccccc cccc cccc}
	\hline
	&&&&&\multicolumn{4}{c}{Case A}&\multicolumn{4}{c}{Case B} \\
	$M_{ini}$	& $M_{fin}$	& $M_{He}$ 	& $M_{CO}$ 	& 	C/O		 & $E_{tot}$ 	&$M_{^{56}Ni}$& log $T_{max}$&	fate	& $E_{tot}$ & $M_{^{56}Ni}$& log $T_{max}$	& fate \\
	\hline
	\multicolumn{3}{c}{Zero metallicity stars} \\
	\hline
	100		&	100.0	&	52.5		&	45.0		&	0.161	&	-		&	-		&	-		&PPISN	&	-		&	-		&	-		&	PPISN	\\
	120		&	120.0	&	57.8		&	56.2		&	0.151	&	-		&	-		&	-		&PPISN	&	-		&	-		&	-		&	PPISN	\\
	125		&	125.0	&	57.0		&	57.0		&	0.141	&	-		&	-		&	-		&PPISN	&	-		&	-		&	-		&	PPISN	\\
	130		&	130.0	&	68.3		&	60.3		&	0.139	&	-		&	-		&	-		&PPISN	&	-		&	-		&	-		&	PPISN	\\
	135		&	135.0	&	70.7		&	62.8		&	0.139	&	-		&	-		&	-		&PPISN	&	-		&	-		&	-		&	PPISN	\\
	140		&	140.0	&	75.2		&	64.9		&	0.130	&	-		&	-		&	-		&PPISN	&	-		&	-		&	-		&	PPISN	\\
	145		&	145.0	&	78.2		&	68.6		&	0.126	&	11.30	&	0.095	&	9.562	&PISN	&	9.46		&	0.011	&	9.531	&	PISN		\\
	150		&	150.0	&	79.9		&	71.3		&	0.125	&	14.05	&	0.154	&	9.571	&PISN	&	12.76	&	0.038	&	9.544	&	PISN		\\
	155		&	155.0	&	82.2		&	73.6		&	0.119	&	16.94	&	0.263	&	9.583	&PISN	&	15.02	&	0.063	&	9.550	&	PISN		\\
	160		&	160.0	&	79.7		&	75.8		&	0.128	&	18.50	&	0.375	&	9.592	&PISN	&	16.51	&	0.086	&	9.556	&	PISN		\\
	180		&	180.0	&	91.1		&	86.7		&	0.121	&	33.46	&	3.778	&	9.642	&PISN	&	30.97	&	0.555	&	9.595	&	PISN		\\
	200		&	200.0	&	98.9		&	94.7		&	0.114	&	42.58	&	9.401	&	9.677	&PISN	&	40.48	&	2.049	&	9.621	&	PISN		\\
	220		&	220.0	&	104.1	&	100.9	&	0.112	&	53.88	&	14.52	&	9.694	&PISN	&	46.69	&	3.384	&	9.632	&	PISN		\\
	240		&	240.0	&	107.9	&	106.8	&	0.105	&	56.08	&	19.85	&	9.717	&PISN	&	55.15	&	5.895	&	9.648	&	PISN		\\
	260		&	260.0	&	121.9	&	119.3	&	0.102	&	81.91	&	42.21	&	9.806	&PISN	&	76.76	&	16.26	&	9.691	&	PISN		\\
	265		&	265.0	&	122.9	&	119.5	&	0.102	&	-		&	-		&	-		&	BH	&	78.82	&	18.07	&	9.699	&	PISN		\\
	270		&	270.0	&	127.1	&	123.2	&	0.100	&	-		&	-		&	-		&	BH	&	85.56	&	22.70	&	9.715	&	PISN		\\
	275		&	275.0	&	128.4	&	124.9	&	0.098	&	-		&	-		&	-		&	BH	&	86.78	&	23.23	&	9.717	&	PISN		\\
	280		&	280.0	&	131.4	&	127.8	&	0.097	&	-		&	-		&	-		&	BH	&	91.12	&	26.19	&	9.726	&	PISN		\\
	300		&	300.0	&	140.3	&	137.2	&	0.093	&	-		&	-		&	-		&	BH	&	109.2	&	45.21	&	9.808	&	PISN		\\
	320		&	320.0	&	145.6	&	142.3	&	0.090	&	-		&	-		&	-		&	BH	&	-		&	-		&	-		&	BH		\\
	\hline
	\multicolumn{3}{c}{1/10 $Z_\odot$ stars}\\
	\hline
	110		&	53.5		&	53.5		&	49.1		&	0.149	&	-		&	-		&	-		&PPISN	&	-		&	-		&	-		&	PPISN	\\
	115		&	57.1		&	57.1		&	49.9		&	0.138	&	-		&	-		&	-		&PPISN	&	-		&	-		&	-		&	PPISN	\\
	120		&	61.1		&	61.1		&	53.4		&	0.120	&	3.02		&	0.001	&	9.521	&PISN	&	2.29		&	0.000	&	9.505	&	PISN		\\
	125		&	61.9		&	61.9		&	54.4		&	0.126	&	3.70		&	0.002	&	9.525	&PISN	&	2.88		&	0.000	&	9.507	&	PISN		\\
	130		&	64.9		&	64.9		&	57.1		&	0.115	&	6.22		&	0.017	&	9.539	&PISN	&	5.02		&	0.000	&	9.517	&	PISN		\\
	140		&	68.2		&	68.2		&	60.6		&	0.148	&	6.88		&	0.017	&	9.540	&PISN	&	5.61		&	0.000	&	9.517	&	PISN		\\
	150		&	71.8		&	71.8		&	67.6		&	0.105	&	11.67	&	0.125	&	9.571	&PISN	&	10.25	&	0.026	&	9.541	&	PISN		\\
	160		&	77.7		&	77.7		&	72.0		&	0.097	&	17.87	&	0.412	&	9.597	&PISN	&	16.19	&	0.101	&	9.560	&	PISN		\\
	180		&	92.7		&	92.7		&	84.6		&	0.122	&	34.10	&	2.829	&	9.636	&PISN	&	31.11	&	0.394	&	9.591	&	PISN		\\
	200		&	101.4	&	101.4	&	95.2		&	0.058	&	47.30	&	15.77	&	9.709	&PISN	&	46.12	&	4.74		&	9.646	&	PISN		\\
	220		&	110.3	&	110.3	&	105.1	&	0.060	&	58.30	&	27.21	&	9.756	&PISN	&	58.79	&	10.34	&	9.674	&	PISN		\\
	240		&	119.1	&	119.1	&	111.4	&	0.074	&	69.03	&	37.51	&	9.795	&PISN	&	70.49	&	14.96	&	9.692	&	PISN		\\
	245		&	120.7	&	120.7	&	112.6	&	0.075	&	71.42	&	39.62	&	9.802	&PISN	&	72.75	&	15.73	&	9.694	&	PISN		\\
	250		&	123.0	&	123.0	&	114.5	&	0.075	&	-		&	-		&	-		&	BH	&	76.18	&	18.47	&	9.703	&	PISN		\\
	255		&	127.2	&	127.2	&	119.5	&	0.069	&	-		&	-		&	-		&	BH	&	82.98	&	23.68	&	9.728	&	PISN		\\
	260		&	129.6	&	129.6	&	121.0	&	0.070	&	-		&	-		&	-		&	BH	&	86.01	&	25.49	&	9.731	&	PISN		\\
	265		&	132.5	&	132.5	&	124.2	&	0.068	&	-		&	-		&	-		&	BH	&	90.80	&	30.43	&	9.752	&	PISN		\\
	270		&	134.4	&	134.4	&	125.3	&	0.069	&	-		&	-		&	-		&	BH	&	94.37	&	33.17	&	9.765	&	PISN		\\
	275		&	136.5	&	136.5	&	127.2	&	0.068	&	-		&	-		&	-		&	BH	&	98.47	&	36.75	&	9.785	&	PISN		\\
	280		&	142.4	&	142.4	&	136.4	&	0.051	&	-		&	-		&	-		&	BH	&	-		&	-		&	-		&	BH		\\
	300		&	151.5	&	151.5	&	143.4	&	0.050	&	-		&	-		&	-		&	BH	&	-		&	-		&	-		&	BH		\\
	320		&	166.7	&	166.7	&	157.3	&	0.038	&	-		&	-		&	-		&	BH	&	-		&	-		&	-		&	BH		\\
	\hline
	\end{tabular}
\end{table*}

\subsection{Common properties of evolution and explosion}
In this subsection, we describe evolution properties that are common among zero and 1/10 $Z_\odot$ metallicities and case A and case B expressions.
For this purpose, zero metallicity stars with case A explosion calculations are taken as examples.

\begin{figure}
\centering
	\includegraphics[width=\columnwidth]{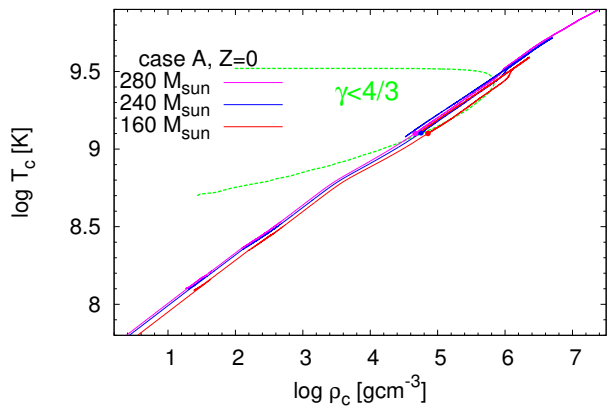}
	\caption{Evolution of the case A zero metallicity models
	having initial masses of 160 M$_\odot$ (red), 240 M$_\odot$ (blue), and 280 M$_\odot$ (Magenta) in a central density-temperature plane.
	The evolution and the hydrodynamic codes are switched at the points of log $T_c$ $\sim$ 9.1 indicated in the figure.
	}
	\label{rhotmp}
\end{figure}
For selected models of the case A zero metallicity set, Figure~\ref{rhotmp} shows evolution of stellar centers in a density-temperature plane.
This figure provides concise information about what kind of reactions affect the evolution of these stars.
The green line shows the boundary of the adiabatic index $\gamma \leq 4/3$.
This reduction of $\gamma$ is due to electron-positron pair creation, which transforms the thermal part of the internal energy to the rest mass, resulting in pressure decrease.
Therefore, as the stellar center enters into the domain, the evolution timescale becomes shorter and shorter.
The core contraction induces temperature increase.
After core carbon burning, neon starts to burn as the central temperature reaches log $T_c \sim 9.3$.
Since neutrino cooling becomes active at that moment, heating by neon burning has a small effect on the evolution, and the core continues to contract.
On the other hand, subsequent oxygen burning more effectively heats the core.
After the ignition of oxygen at log $T_c$ $\sim$ 9.5, the two less massive models explode, while the most massive 280 M$_\odot$ model collapses.

\begin{figure}
\centering
	\includegraphics[width=\columnwidth]{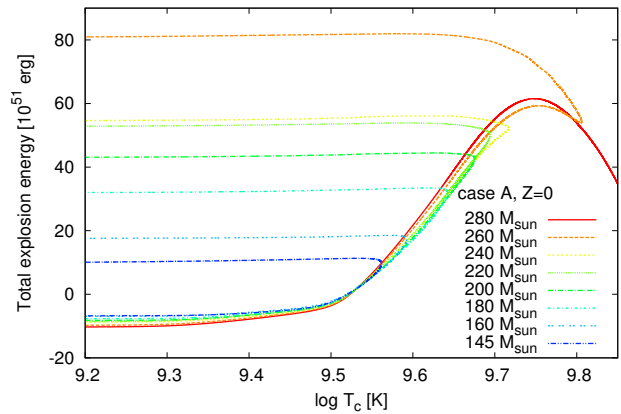}
	\caption{Evolution of the total energy during the explosion.
	Selected models are 145, 160, 180, 200, 220, 240, 260, and 280 M$_\odot$ case A zero metallicity models.
	}
	\label{energy_z0wm}
\end{figure}
In Figure~\ref{energy_z0wm}, the evolution of the non-relativistic total energy as a function of the central temperature is shown for selected case A zero metallicity models.
The total energy is defined as $\int ( \frac{1}{2}U^2 + e^{therm+pair} - \frac{GM_b}{r} ){\rm d}M_b$, where $U$ is the radial velocity of the gas.
This figure clearly shows that two reactions are responsible for determination of the fate of a contracting very massive star.
The first reaction is oxygen burning, which ignites when the central temperature becomes log $T_c=9.5$.
If oxygen burning supplies enough energy to invert the core motion, the core starts to expand, resulting in an explosion as a PISN.
However, when the central temperature reaches log $T_c=9.75$, the total energy starts to decrease.
This is due to photo dissociation reaction.
This reaction, similar to the pair creation reaction, transforms the thermal energy to the rest mass and reduces the pressure.
If destabilization by central photo dissociation overcomes the energy inputs by surrounding oxygen burning, the star collapses and forms a black hole.
Therefore, change of the fate can be seen through the difference among the maximum temperatures reached during the explosion for different cores.
Our calculation shows that log $T_c = 9.8$ is the maximum temperature for a CO core to invert the motion.

\begin{figure}
	\includegraphics[width=\columnwidth]{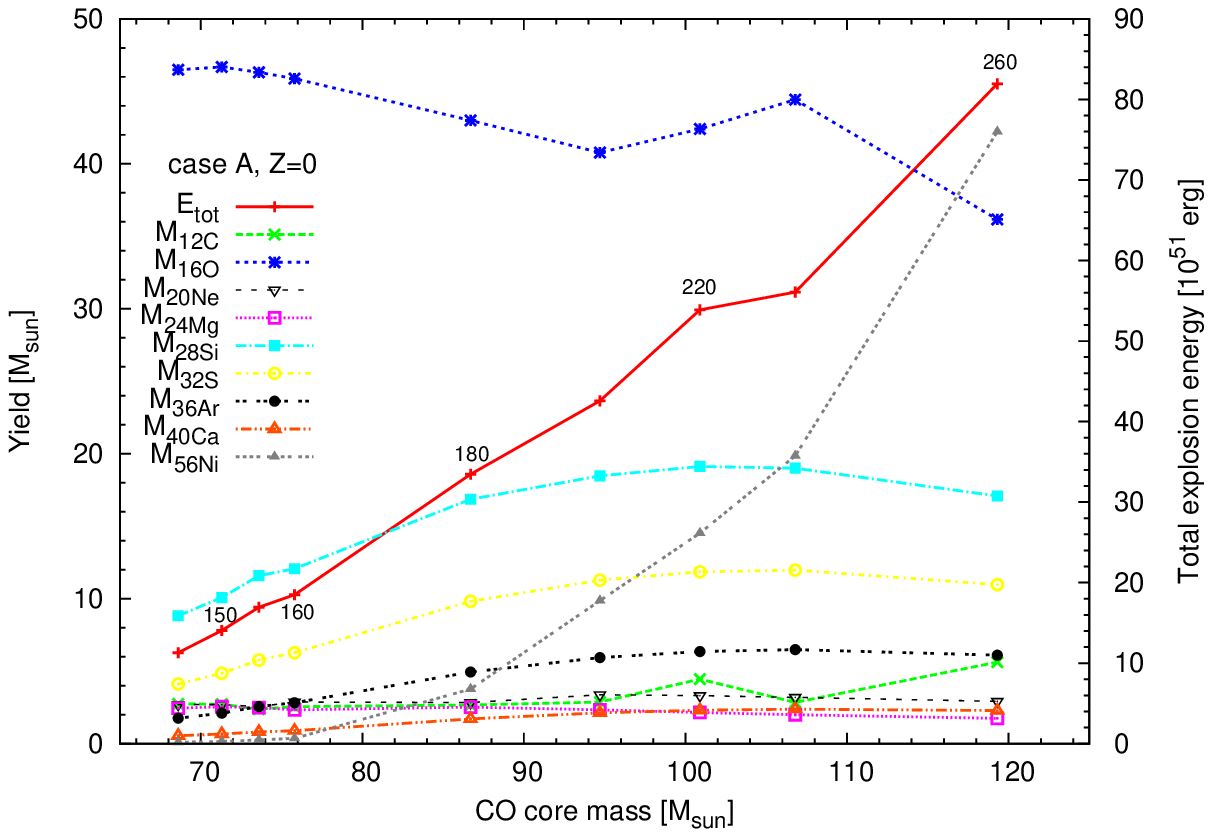}
	\caption{Yields and the total explosion energy as functions of the CO core mass.
	All zero metallicity models that explode with the case A energy generation rates are plotted,
	they are 145, 150, 155, 160, 180, 200, 220, 240, and 260 M$_\odot$ models.
	Numbers indicated near the total energy show the corresponding initial masses.}
	\label{yield_z0}
\end{figure}
Explosion properties are well correlated with the CO core mass, rather than the initial mass.
For case A zero metallicity explosions, the explosion energy as well as the yields of representative isotopes
are summarized as functions of the CO core mass in Figure~\ref{yield_z0}.
The explosion energy is roughly proportional to the CO core mass.
On the other hand, the yields of isotopes have more complicated dependence.
Isotopes are divided into three groups.
The first group is hydrostatic burning products: lighter elements than magnesium.
Their yields weakly depend on the CO core mass.
The second group is oxygen burning products, which consist of isotopes heavier than silicon and lighter than calcium.
Their yields have a peak at the intermediately massive CO core of $\sim$100 $M_\odot$.
This is because a less massive CO core yields less massive oxygen burning products,
while a certain amount of oxygen burning products are processed in a more massive core.
The ratio of these yields are almost independent from the CO core mass.
The last group is silicon burning products, and most of them are dominated by $^{56}$Ni.
For less massive models, almost no $^{56}$Ni is produced by the explosion,
since the maximum temperatures during the explosion are too small to burn oxygen burning products.
Massive models, contrastingly, yield large amount of $^{56}$Ni,
because the major part of the material in the high temperature central regions are transformed into $^{56}$Ni.
Hence, silicon burning products show a strong dependence on the CO core mass.

\subsection{Comparison between stars with the two metallicities}

Stars with the two different metallicities develope different envelope structures.
Zero metallicity stars retain their envelopes during the evolution, while 1/10 Z$_\odot$ stars completely lose them.
Table~\ref{stars} shows that this difference affects the fate of less massive stars of 120-140 M$_\odot$.

With a fixed initial mass, CO core masses and thus explosion energies are similar for stars with the two metallicities.
A star with 1/10 Z$_\odot$ is much easier to explode, because nothing interferes with the expansion.
In contrast, a zero metallicity star fails to explode with the same explosion energy.
For these stars, core expansion forms a shock at the core surface.
When the shock passes the core/envelope boundary, the momentum and the kinetic energy of the core is consumed to accelerate the envelope.
As a result, the core stops expanding and falls back into the center.
Resulting mass ranges of PISN are,
for case A calculations, $M_{ini} \in [145, 260]$ M$_\odot$ for zero metallicity stars and it shifts to $M_{ini} \in [120, 245]$ M$_\odot$ for Z=1/10 Z$_\odot$ stars.
For case B calculations, $M_{ini}  \in [145, 300]$ M$_\odot$ for zero metallicity and $M_{ini} \in [120, 275]$ M$_\odot$ for 1/10 Z$_\odot$.

\subsection{Difference due to adopting different energy generation rates}

\begin{figure*}
\centering
	\includegraphics[width=1.8\columnwidth]{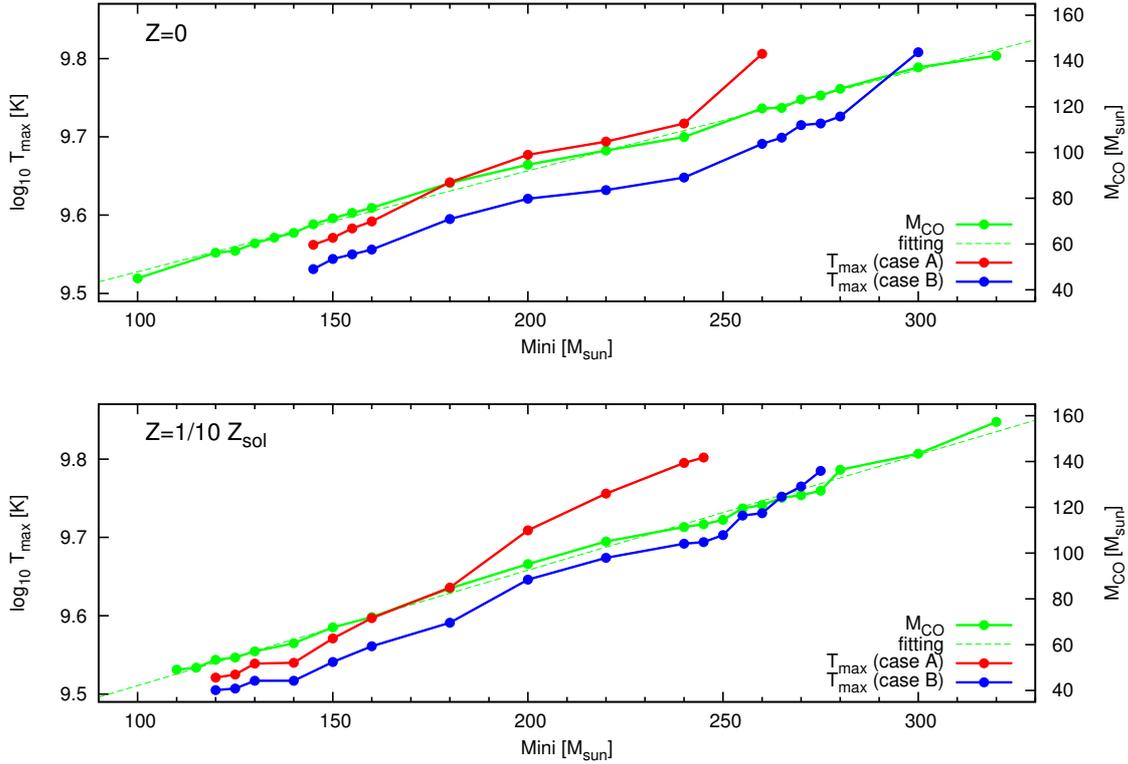}
	\caption{The maximum temperature reached during the explosion and the CO core mass as functions of the initial mass for all exploded models.
	Top panel shows zero metallicity results and bottom shows 1/10 Z$_\odot$ results.
	The green dotted line in each panel is a fitting function for the CO core mass.}
	\label{range}
\end{figure*}
By neglecting the thermal part in the chemical potential, the approximate entropy expression applied to the case B calculations overestimates the nuclear energy generation rate.
In Figure~\ref{range}, the maximum central temperature reached during the explosion, as well as the CO core mass, are shown as functions of the initial mass.
This figure shows all case B models explode with lower central temperatures than the case A counterparts.
The central temperature indicates how much amount of oxygen is consumed in the contracting CO core,
in other words, the higher the central temperature is, the larger amount of oxygen are consumed.
Hence, using the more efficient energy generation rate, the case B star explodes with a smaller amount of oxygen burned.

\begin{figure*}
	\includegraphics[width=1.8\columnwidth]{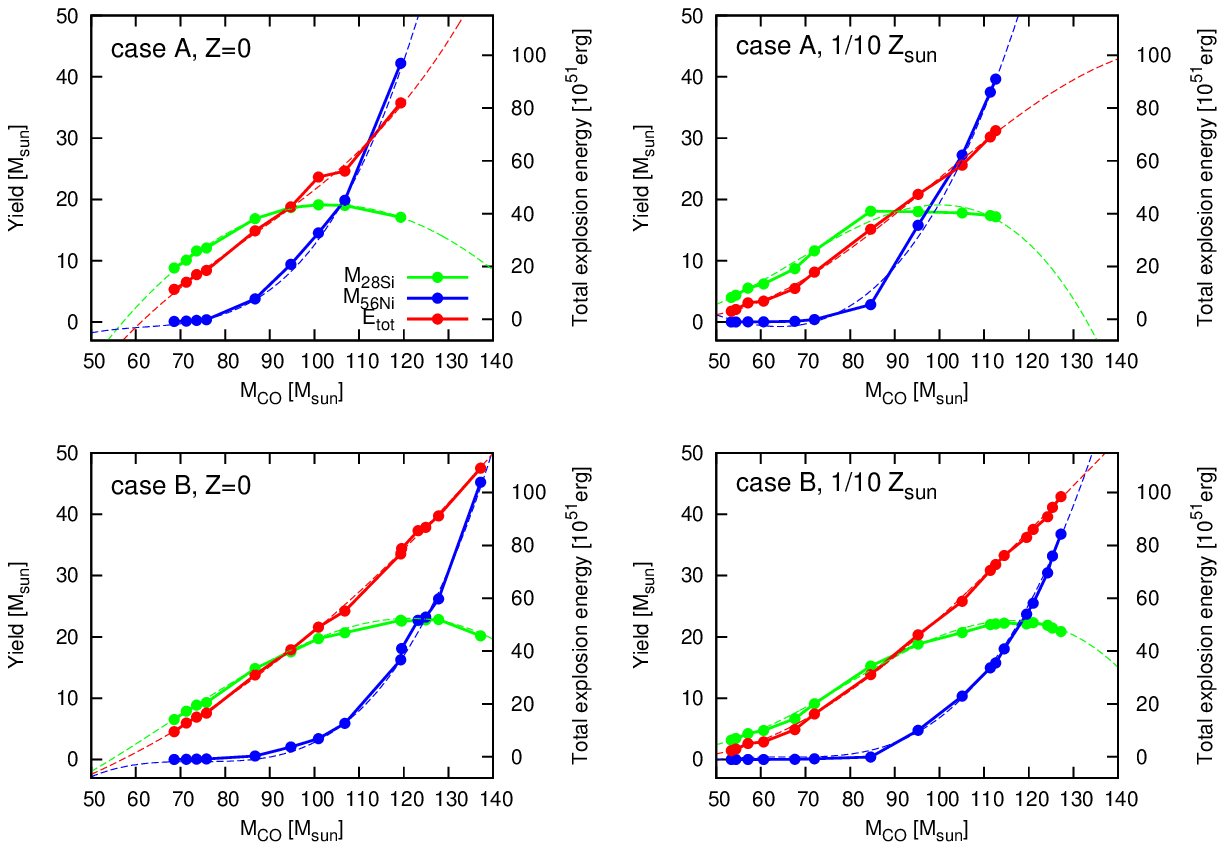}
	\caption{Yields of $^{28}$Si and $^{56}$Ni and the total explosion energy as functions of the CO core mass for all exploded models.
	Top panels show results of case A calculations and bottoms show that of case B.
	Left panels are zero metallicity results and right panels are 1/10 Z$_\odot$ results.
	Dotted lines show the fitting polynomials.}
	\label{fit}
\end{figure*}
As already discussed in \S 4.1, most massive stars that explode as a PISN show a universal maximum temperature of log $T_{max}=9.8$.
Since the maximum temperature of the case B calculation is lower than that of the case A, 
some massive stars that collapse with the exact energy generation rate become able to explode with the approximate energy generation rate.
Left panels of Figure~\ref{fit} shows that the maximum CO core mass for PISN extends from 119 M$_\odot$ to 137 M$_\odot$ for zero metallicity models.
Similarly, the extension for 1/10 Z$_\odot$ models is from 112 M$_\odot$ to 127 M$_\odot$ (the right panels).
These figures also show that the explosion energy is roughly proportional to the CO core mass,
and the dependence is almost identical in the four sets of calculations.
Therefore, case B calculations result in the larger maximum explosion energy of PISN than case A.
On the other hand, since the $^{56}$Ni yield strongly depends on the maximum temperature,
the most massive models for each set of calculations yield similarly largest amount of $^{56}$Ni.

\begin{figure}
	\includegraphics[width=\columnwidth]{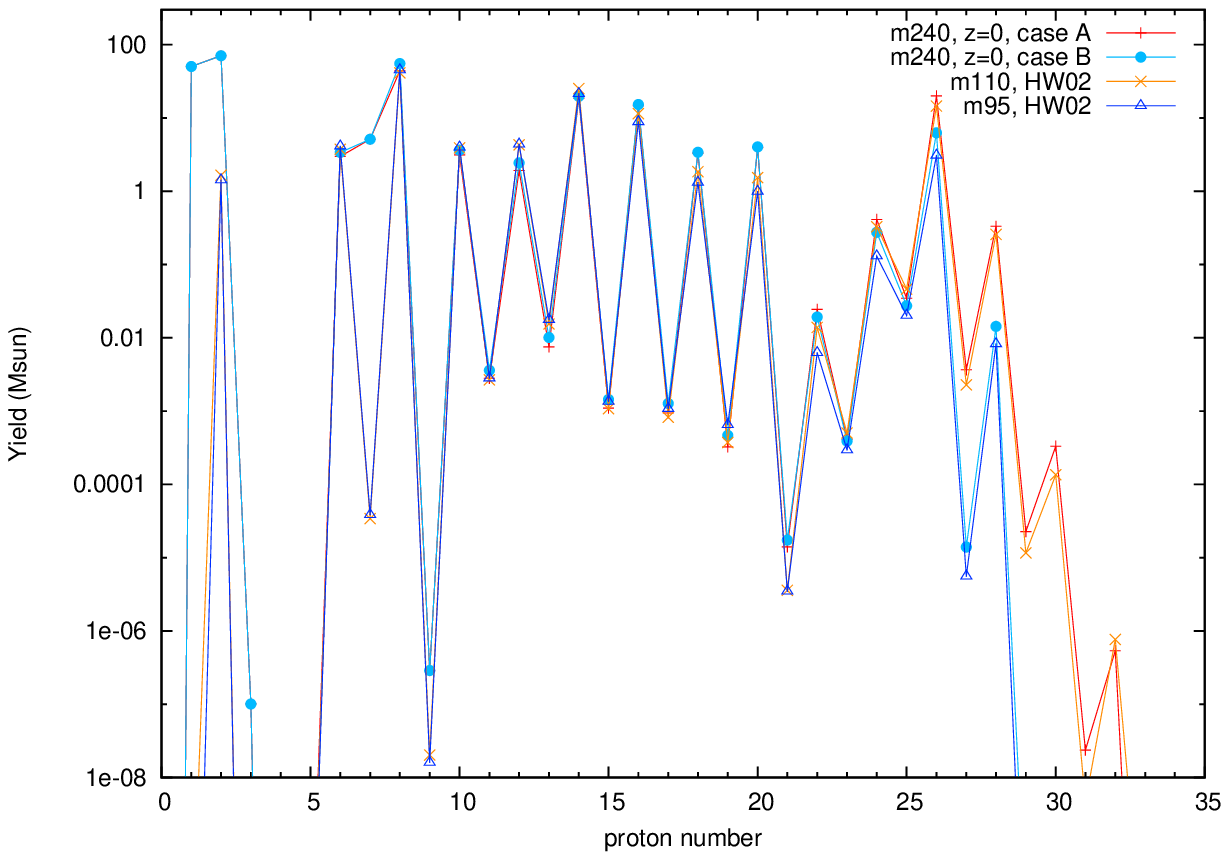}
	\caption{
	Composition patterns of PISN yields of different models.
	Red-plus patterns and cyan-point patterns show yields of case A and case B zero metallicity 240 M$_\odot$ calculations, respectively.
	Similarly, patterns shown by orange-cross and blue-triangle are yields of 110 M$_\odot$ and 95 M$_\odot$ helium star models taken by \citet{heger&woosley02}.	
	}
	\label{yield}
\end{figure}
In Figure~\ref{yield}, yield comparison among four models, two from our calculation and the other two from \citet{heger&woosley02}, are shown.
With the exact entropy expression, yield of our zero metallicity 240 M$_\odot$ model shows almost the same composition pattern as the 110 M$_\odot$ helium star model by \citet{heger&woosley02}.
As both models have almost the same He core mass of $\sim$110 M$_\odot$, this agreement indicates the physical consistency of the two calculations.
The only difference between the two is production of hydrogen, helium, lithium, and nitrogen in our model.
These elements are produced at the base of the hydrogen envelope, which is omitted in the model by \citet{heger&woosley02}.
If the approximate entropy expression is applied to the same 240 M$_\odot$ star instead, the composition pattern is altered.
Reflecting the lower temperature during the explosion at the central region, the case B 240 M$_\odot$ model yields smaller amount of heavy elements heavier than nickel.
The composition pattern resembles the pattern of the 95 M$_\odot$ helium star model by \citet{heger&woosley02},
however, the helium star mass is much smaller than the core mass of the 240 M$_\odot$ model.

\begin{figure}
	\includegraphics[width=\columnwidth]{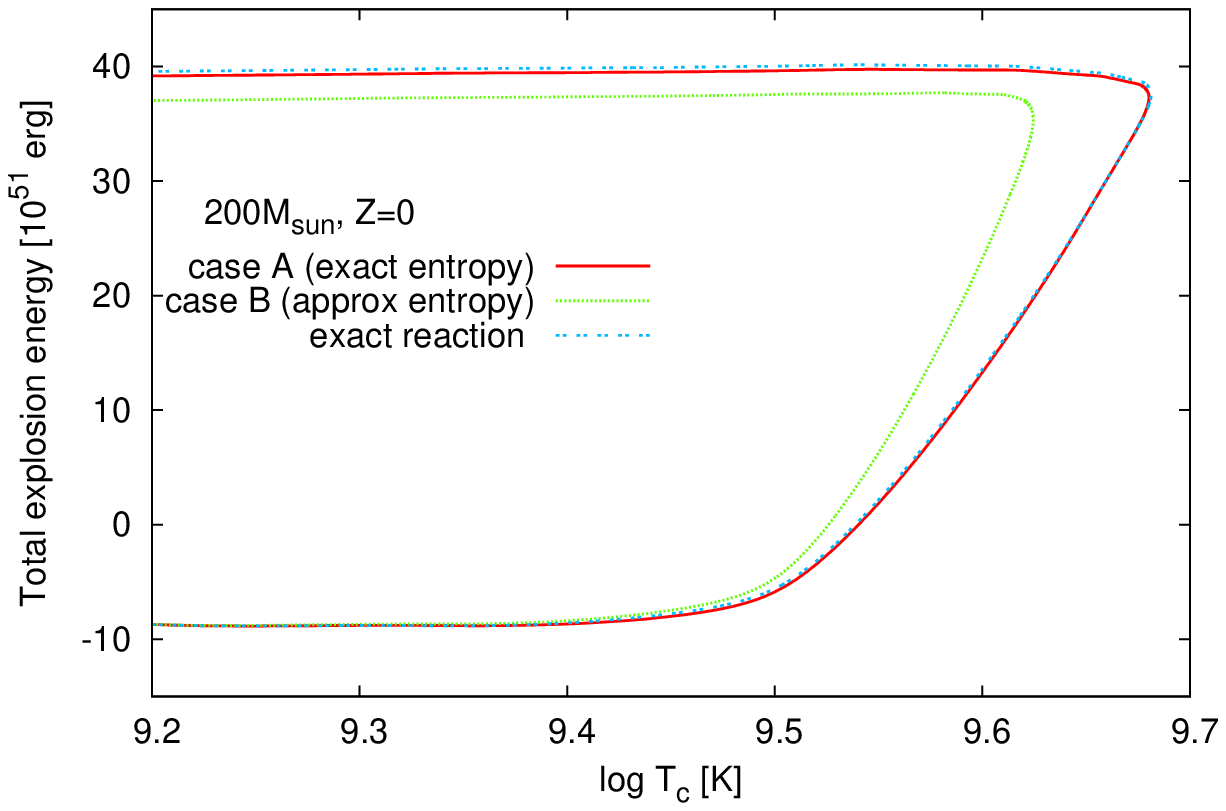}
	\caption{
	Evolution of the total energy during the explosion.
	All models are zero metallicity 200 M$_\odot$ stars, but applied energy generation rates are different;
	the exact entropy expression (red, solid),
	the approximate entropy expression (green, dotted),
	and the exact reaction expression (blue, dash-dotted).
	}
	\label{energy_comp}
\end{figure}
Finally, we show the comparative result of zero metallicity 200 M$_\odot$ explosions, for which three different energy generation rates are adopted (Figure~\ref{energy_comp}).
Obviously, two explosions calculated with exact expressions almost perfectly coincide with each other, though the equations actually treated are different.
This provides a definite confirmation of our perspective.
On the other hand, the model with the approximate entropy expression shows steeper increase of the total energy,
showing that the nuclear energy generation rate is overestimated with the approximate expression.

\section{Impact on the RATE ESTIMATE}

So far, we have demonstrated that explosion calculation with an approximate energy generation rate, which might be widely used in the community,
fails to reproduce correct properties of PISN, such as the too large maximum explosion energy and the too large maximum mass for PISN.
Since these parameters are crucial for the theoretical prediction of observability of high-redshift PISNe,
it is important to see to what extent the calculated rate is affected by adopting the results.

\begin{table}
	\centering
	\caption{Constants used to fit the CO core mass.}
	\label{coeff_co}
	\begin{tabular}{c cc}
	\hline
	metallicity	&$a_{0,CO}$&$a_{1,CO}$	\\
	\hline
	Zero			&  3.781 & 0.441 \\
	1/10 Z$_\odot$	& -8.280 & 0.504 \\
	\hline
	\end{tabular}
\end{table}

\begin{table*}
	\centering
	\caption{Constants that is used to fit the $^{56}$Ni yield and the total explosion energy.}
	\label{coeff_q}
	\begin{tabular}{c cccc cccc}
	\hline
	expression	&	$a_{0,Ni}$&$a_{1,Ni}$&$a_{2,Ni}$&$a_{3,Ni}$	&	$a_{0,E}$&$a_{1,E}$&$a_{2,E}$&$a_{3,E}$\\
	\hline
	\multicolumn{3}{c}{Zero metallicity}\\
	\hline
	case A		&	-5.24E1	&	2.40		&	-3.80E-2	&	2.05E-4	&	2.22E2	&	6.53		&	-6.27E-2	&	2.45E-4\\
	case B		&	-7.22E1	&	2.91		&	-3.93E-2	&	1.77E-4	&	-2.96E1	&	1.58E-1	&	5.88E-3	&	2.61E-6\\
	\hline
	\multicolumn{3}{c}{1/10 $Z_\odot$}\\
	\hline
	case A		& 1.52E1	&-1.79E-2	&-1.11E-2	& 1.17E-4	& 6.15E1	&-3.07	& 4.51E-2	&-1.52E-4	\\
	case B		&-40.2E1	& 1.88	&-2.90E-2	& 1.49E-4	& 3.73E1	&-1.97	& 2.86E-2	&-7.37E-5	\\
	\hline
	\end{tabular}
\end{table*}
Two mechanisms that illuminate PISNe are considered:
the first one is a radioactive decay of $^{56}$Ni and the other one is interaction with the optically thick circumstellar medium (CSM).
For the two mechanisms, either the ejected mass of $^{56}$Ni or the total explosion energy is used as an indicator of the PISN luminosity.
As already discussed, these explosion properties are well correlated with the CO core mass of the star.
The mass of the CO core is in good linear correlation with the initial mass for both zero metallicity and 1/10 Z$_\odot$ models.
This can be expressed as
\begin{eqnarray*}
	M_{CO} = a_{0,CO} + a_{1,CO} \times M_{ini},
\end{eqnarray*}
where constant coefficients are shown in Table~\ref{coeff_co} (see Figure~\ref{range} for the fitting).
Then, the $^{56}$Ni yield and the explosion energy are fitted by cubic equations of the CO core mass as
\begin{eqnarray*}
	M_{^{56}Ni} &=& a_{0,Ni} + a_{1,Ni}  \times M_{CO} + a_{2,Ni}  \times M_{CO}^2 + a_{3,Ni}  \times M_{CO}^3 \\
	E_{tot} &=& a_{0,E}   + a_{1,E}   \times M_{CO} + a_{2,E}   \times M_{CO}^2 + a_{3,E}    \times M_{CO}^3,
\end{eqnarray*}
where constant coefficients are listed in Table~\ref{coeff_q} (Figure~\ref{fit}).

In the following subsections, we assume two different initial mass functions (IMFs) having different indices on the initial mass.
For an IMF, a simple form having an index $\alpha$ is used,
\begin{eqnarray*}
	F_\alpha(M_{ini})dM_{ini} = A M_{ini}^{-\alpha} dM_{ini},
\end{eqnarray*}
where $A$ is a normalization constant.
Rate estimates of relatively low-redshift PISNe are done with the Salpeter value, $\alpha=2.35$, and a flat value $\alpha=0$ is applied for high-redshift estimates.
The latter represents the IMF of zero metallicity stars for the considered mass range better than the Salpeter value \citep{hirano+14}.
Note that presented models of zero- and 1/10 Z$_\odot$ are simply regarded as models without and with mass loss in this analysis.
This is why these two IMFs are applied for both zero- and 1/10 Z$_\odot$ metallicity models,
even though stellar metallicity would be related to the index of the IMF in reality.

The number of PISNe is integrated with the fitting functions.
For this purpose, a condition function $f_{con}$, which takes one for stars that meet the condition, otherwise becomes zero, is defined for each condition.
The integral
\begin{eqnarray*}
	q_{con} = \frac	{ \int_{M_{PISN, min}}^{M_{PISN, max}} f_{con}(M') \times F_\alpha(M') dM' }
				{ \int_{10 \mathrm{M}_{\odot} }^{20 \mathrm{M}_{\odot} } F_\alpha(M') dM' }
\end{eqnarray*}
is numerically summed up, where $M_{PISN, min}$ and $M_{PISN, max}$ are the minimum and the maximum initial masses for PISNe.
In order to fix the normalization, we use the number of stars within the mass range of $M_{ini} \in [10,20]$ M$_\odot$,
which may provide the number of type II-P core collapse supernovae at that frame.
Then, the $q_{con}$ shows the number fraction of PISNe, the property of which meets the imposed condition.

\begin{figure*}
	\includegraphics[width=1.8\columnwidth]{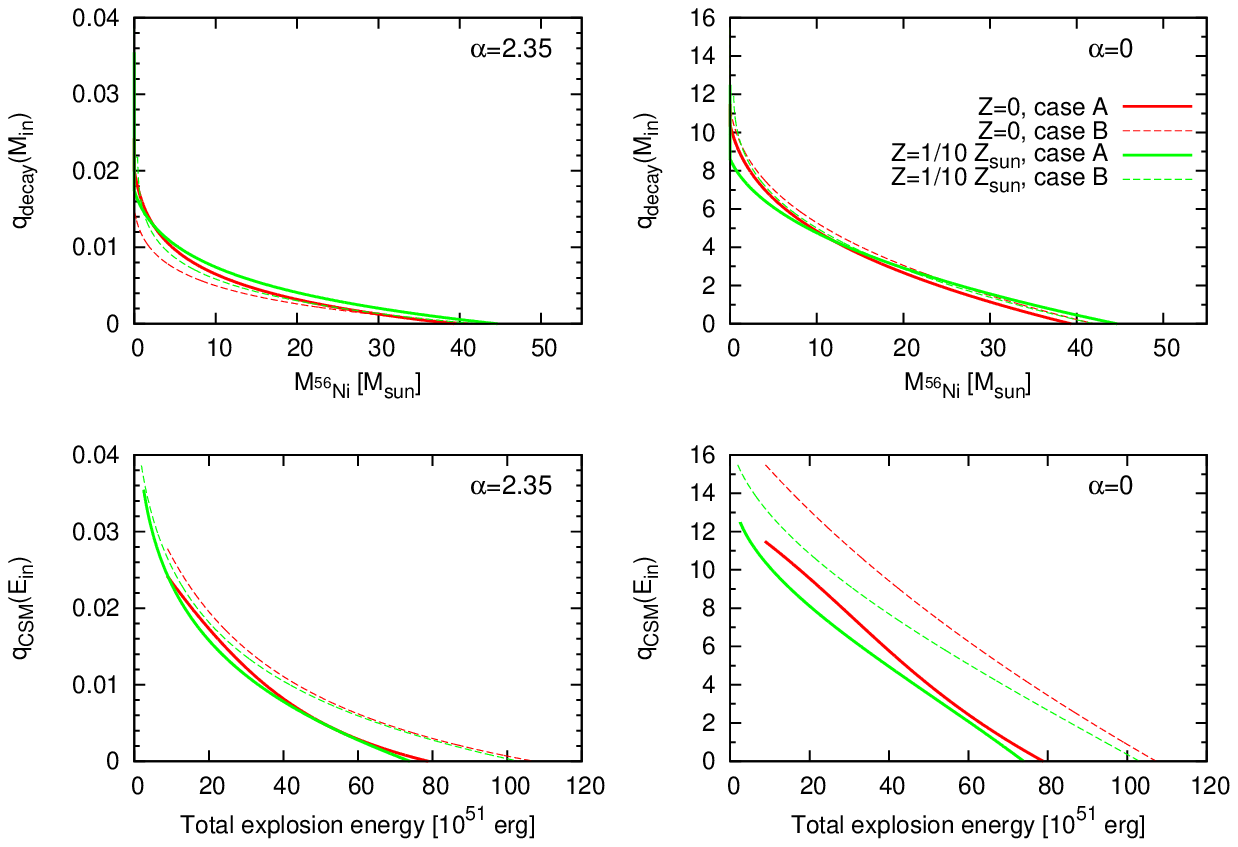}
	\caption{Cumulative number fractions calculated for the $^{56}$Ni yield (top panels) and for the total explosion energy (bottom panels).
	Left panels are calculated with $\alpha$=2.35 IMF, while $\alpha=0$ is adopted to the right.
	In each panel, every four sets of calculations are shown (case A zero metallicity as red bold, case B zero metallicity as red dotted,
	case A 1/10 Z$_\odot$ as green bold, and case B 1/10 Z$_\odot$ as green dotted).}
	\label{intg}
\end{figure*}

\subsection{Radioactive decay}

A large amount of $^{56}$Ni synthesized by a massive PISN explosion gradually decays, emitting gamma rays and heating up the surroundings.
This energy is converted into thermal radiation and accounts for the luminosity of a supernova.
As a crude estimate, a simple linear correlation between $^{56}$Ni and the supernova luminosity is assumed.
With a condition function of
\begin{eqnarray*}
	f_{decay}(M_{in}; M') =	\left\{ \begin{array}{ll}
					1	&	(M_{ ^{56}Ni }(M') > M_{in}) \\
					0	&	(otherwise),
 				 \end{array} \right.
\end{eqnarray*}
the cumulative number fraction is integrated as
\begin{eqnarray*}
	q_{decay}(M_{in}) = \frac	{ \int_{M_{PISN, min}}^{M_{PISN, max}} f_{decay}(M_{in}; M') \times F_\alpha(M') dM' }
						{  \int_{10 \mathrm{M}_{\odot} }^{20 \mathrm{M}_{\odot} } F_\alpha(M') dM' }.
\end{eqnarray*}
The integral, $q_{decay}(M_{in})$, shows the number of PISNe that yield a larger amount of $^{56}$Ni than $M_{in}$ per one CCSN.

Upper panels of Figure~\ref{intg} show the cumulative number fractions as functions of $^{56}$Ni yield,
and the left panel shows the result of $\alpha=2.35$ and the right panel is for the $\alpha=0$ result.
For $\alpha=2.35$ integrations, case B models predict a smaller number of PISNe than the case A models.
Meanwhile, number fractions are larger or similar with the flat IMF of $\alpha=0$.
This is because the case B models require a more massive initial mass than the case A to yield the same amount of $^{56}$Ni,
and massive stars are rarer with the index of $\alpha=2.35$, while the flat IMF allows a large number of such massive stars to exist.

SN 2007bi is a famous superluminous supernova (SLSN) which has a peak absolute luminosity of -21.35 mag \citep{gal-yam+09}.
Using the peak luminosity and the late-time decay tail, $\sim$5 M$_\odot$ of $^{56}$Ni is estimated to be ejected by the explosion.
As well as the large $^{56}$Ni mass, the large total ejecta mass of $\sim$100 M$_\odot$
and the large kinetic energy of $\sim$$10^{53}$ erg are indicated by adopting scaling relations of \citet{arnett82}.
PISN model has given a possible explanation for these characteristics \citep[and references therein]{gal-yam12}.
Taking this as an example, we compare the number fractions of 1/10 Z$_\odot$ PISNe that yield more $^{56}$Ni than 5 M$_\odot$, $q_{decay}(M_{in}=5\mathrm{M}_\odot)$.
The required minimum initial mass becomes 184 M$_\odot$ for 1/10 Z$_\odot$ case A models, but it increases to 207 M$_\odot$ for case B.
The cumulative number fractions are $1.03\times10^{-2}$ and $8.63\times10^{-3}$ for $\alpha$=2.35, respectively.
Therefore, with the approximate energy generation rate, the detectable number of PISNe is underestimated to be 79\% of the fiducial value for the Salpeter IMF.
With the flat IMF, the cumulative number fractions are 6.08 and 6.72.

\subsection{CSM interaction}
Another mechanism that possibly illuminate a PISN is collision with optically thick CSM, by which the SN ejecta and the CSM are heated and emit thermal radiation.
A high conversion efficiency of $\sim$10\% from the kinetic to the thermal energy is often assumed in this process,
so that this mechanism can explain the high luminosities of some observed SLSNe \citep[e.g.,][]{moriya+10}.
In this analysis, the total energy of the explosion $E_{tot}$ is taken as the condition variable,
and the cumulative number fraction of PISNe is calculated as
\begin{eqnarray*}
	q_{CSM}(E_{in}) &=& \frac	{ \int_{M_{PISN, min}}^{M_{PISN, max}} f_{CSM}(E_{in}; M') \times F_\alpha(M') dM' }
							{  \int_{10 \mathrm{M}_{\odot} }^{20 \mathrm{M}_{\odot} } F_\alpha(M') dM' }\\
	f_{CSM}(E_{in}; M') &=&	\left\{ \begin{array}{ll}
					1	&	(E_{tot}(M') > E_{in}) \\
					0	&	(otherwise).
 				 \end{array} \right.
\end{eqnarray*}
Since the CSM interaction mechanism requires dense CSM surrounding the SN progenitor, significant mass loss during the lifetime would be needed as we consider a single progenitor.
This means that the 1/10 Z$_\odot$ models that suffer from heavy mass loss provide more consistent result, though we conduct the same analysis to the zero metallicity calculations.

The lower left panel of Figure~\ref{intg}, showing the result of $\alpha$=2.35, shows that the case B estimate gives larger number fraction than the case A result for all $E_{tot}$.
This is due to the larger maximum explosion energy, or the larger maximum mass, of the case B calculations.
Massive PISNe which only explode with the case B energy generation rates are always summed up in the integration, increasing the estimated number fraction.
Thus, the massive population that only exists in the case B calculations artificially increases the observational rates.
This discrepancy between the two cases of energy generation rates is enhanced with the $\alpha$=0 IMF, as shown by the lower right panel of Figure~\ref{intg}.
This is because the massive populations are much larger for the flat IMF than $\alpha$=2.35.

\subsection{Difference between the two metallicity calculations}

Variation in calculations due to different applied metallicities, or different assumptions of mass loss,
is more essential than deviation caused by different energy generation rates.
In an actual estimate of the detection rate of PISN, therefore,
difference between the two metallicity calculations will be considered as the uncertainty in the theory.

As discussed above, the most significant difference between the explosions of two different metallicities is the minimum mass for the PISN explosion.
The less massive stars, which only explode with the efficient mass loss of 1/10 $Z_\odot$, yield almost no $^{56}$Ni.
Therefore, only a minor difference is caused in the detection number of PISN that is illuminated by radioactive decay.
In contrast, since the explosion energy of these less massive stars are substantially small,
these explosions affect the detection number of the CSM interacting PISN by decreasing the lowest detected explosion energy.
In the present work, calculations with only two extreme cases of mass loss assumptions are considered.
Hence, it will be important to be investigated the consequence of the PISN mass range of by adopting moderate mass loss rates.

Moreover, this analysis neglects the contribution from PPISNe that arise from smaller mass range than the PISN progenitors.
Pulses of a PPISN may plurally eject outer parts of the star,
and collisions among these ejecta are suggested to be observed as a luminous transient similar to the CSM interacting SLSN \citep{woosley+07}.
A loosely bound envelope will be much applicable for the ejection than the surface region of a stripped star.
Therefore, contribution from PPISN will also depend on the metallicity of the star, by which the envelope structure is affected.
This will also be important topic to be investigated.

\section{CONCLUSION}

The explosion mechanism of PISN is rather simple, however, it is not trivial how to solve the energetics in actual simulation.
Starting from energy conservation in hydrodynamics, we have derived four exact and two approximate expressions of the stellar energy equation.

Adopting an exact and another approximate expressions, evolution and succeeding explosion of zero- and 1/10 Z$_\odot$ very massive stars are calculated.
Common features are:
\begin{enumerate}
	\item neon burning starts when the central temperature becomes log $T_c$ = 9.3, but this does not significantly affect the evolution,
	\item oxygen burning that ignites at log $T_c$ = 9.5 more effectively heats the core, supplying energy to reverse the motion,
	\item if the central temperature reaches log $T_c$ = 9.8, the core collapses, otherwise it explodes,
	\item PISN yields are divided into three groups of hydrostatic burning products, oxygen burning products, and silicon burning products,
		which have different dependence on the CO core mass, and
	\item the explosion energy is correlated with the CO core mass, approximately in linear.
\end{enumerate}

The difference between the two metallicity models is the minimum initial mass to become a PISN.
It is affected by the existence of a hydrogen envelope,
and the value decreases from 145 M$_\odot$ for zero metallicity models with hydrogen envelope to 120 M$_\odot$ for 1/10 Z$_\odot$ fully stripped models.
Different energy generation rates result in different maximum initial masses for PISNe.
The energy generation rate of the approximate entropy expression, which is applied to the case B calculations,
more effectively heats surroundings than the exact expression applied to the case A.
Hence, the maximum initial mass and the explosion energy become larger for the case B than for the case A.
As for $^{56}$Ni yield, since the most massive models yield similar amounts of $^{56}$Ni,
the $^{56}$Ni yield as a function of the CO core mass shifts to higher masses for the case B calculations.

In order to discuss the observational consequences, number fractions of PISNe that meet a criterion are calculated with two different IMFs.
For the criterion, the $^{56}$Ni yield or the total explosion energy is used, which corresponds to a different energy source for the SN luminosity.
Because of the large maximum initial mass and the large maximum explosion energy, application of the approximate rates significantly changes the results.
When the Salpeter IMF is considered, the number fraction is underestimated if radioactive decay is assumed as the energy source,
on the other hand, it is overestimated if the CSM interaction mechanism is assumed.
For the CSM interacting PISN, the overestimation is enhanced with the flat IMF, which has a much larger population of the massive PISNe.

We conclude that, in order to accurately consider the energetics of reactions,
the definition of energy generation rates in the hydrodynamic equation is fundamentally important.
In the context of stellar physics, the energy generation rates are divided into $\epsilon_{grv}$ and $\epsilon_{nuc}$,
however, these terms are related to each other.
Four expressions of the energy conservation presented in this paper are physically equivalent,
and the entropy expression is suitable for a simulation when some reactions are in reaction equilibrium,
as with the case of the pair creation-annihilation reaction in PISNe.


\section*{Acknowledgements}
The authors appreciate to Chris Fryer, the referee of this paper, for valuable comments and suggestions.
The authors are grateful to Stanford Woosley, Alexander Heger, Bill Paxton, Frank Timmes, Norbert Langer, Georges Meynet, and Alessandro Chieffi
for providing detail information about energy generation rates in their code.
The authors also thank to Alex Heger for providing pre-collapse structures of their PISN models.
KT is supported by research fellowships of JSPS for Young Scientists.
KT also acknowledges the usage of the supercomputer at YITP in Kyoto University.
This work is supported in part by the Large Scale Simulation Program (No.14/15-17) of High Energy Accelerator Research Organization (KEK).
This work is also supported in part by the Grant-in-Aid for Scientific Research (24103006, 24105008, 24244028, 24244036, 26104006, 26400271, 15K05093).


\bibliographystyle{mnras}
\bibliography{biblio}

\appendix
\appendix
\section{Definitions of mass densities, number densities, and mole fractions}\label{app1}
When reactions occur, rest mass density becomes a non-conserved variable.
Instead of the rest mass density, one may define the baryon number density $n_b$ as the new conserved variable,
employing the baryon number conservation as the new conservation law.
Using the baryon number density, pseudo mass density, or also called as the baryon mass density,
can be defined as $\rho_b \equiv m_u n_b$, which also conserves regardless of reactions.

The baryon number density is related with the number density of ion, which has the mass number of $A_i$, as
\begin{eqnarray}
	n_b = \sum_{ion} A_i n_i.
\end{eqnarray}
Ionic mole fraction $Y_i$ and mass fraction $X_i$ are defined as
\begin{eqnarray}
	Y_i &\equiv& n_i / n_b \\
	X_i &\equiv& \rho_i / \rho_b,
\end{eqnarray}
where $\rho_i \equiv A_i m_u n_i$ is pseudo mass density of $i$-th ion.
Thus, a relation
\begin{eqnarray}
	Y_i &=& X_i / A_i
\end{eqnarray}
holds.
Using the conservation relation $\sum_{ion} X_i = 1$, one may use the mass fractions as independent variables for chemical composition.

One may define the net electron number density and the net electron mole fraction as
\begin{eqnarray}
	n_e &\equiv& \sum_{ion} Z_i n_i \\
	Y_e &\equiv& n_e/n_b \\
	         & = & \sum_{ion} Z_i Y_i,
\end{eqnarray}
where $Z_i$ is charge number of $i$-th ion.
Via charge neutrality, the net electron number density is related with number densities of electron $n_{e^-}$ and positron $n_{e^+}$ as
\begin{eqnarray}
	n_e = n_{e^-} - n_{e^+}.
\end{eqnarray}
Like an ionic mole fraction, electron and positron mole fractions are defined as
\begin{eqnarray}
	Y_{e^-} &\equiv& n_{e^-} / n_b, \\
	Y_{e^+} &\equiv& n_{e^+} / n_b.
\end{eqnarray}
Then, the relation
\begin{eqnarray}
	Y_e	& = &Y_{e^-} - Y_{e^+}
\end{eqnarray}
is obtained.

Rest mass density of gas composed of photon, ions, electron, and positron are written as
\begin{eqnarray}
	\rho = \sum_{ion} m_i n_i + m_{e^-} n_{e^-} + m_{e^+} n_{e^+}. \label{rest_mass}
\end{eqnarray}
Equating with above relations, one may obtain
\begin{eqnarray}
	\frac{ \rho c^2 }{\rho_b } = \frac{1}{m_u} \Bigl[ \sum_{ion} m_i c^2 Y_i + m_e c^2 Y_e + 2m_e c^2 Y_{e^+} \Bigl], \label{rest_mass2}
\end{eqnarray}
where the relation $m_e \equiv m_{e^-}=m_{e^+}$ is used.
In the right hand side, a reaction only changes the mole fractions,
and thus the equation gives a simple way to calculate the change of the rest mass per baryon by reactions.

\section{The second law of thermodynamics}\label{app2}
Macroscopic expression of the second law of thermodynamics is
\begin{eqnarray}
	{\rm d}E^{rel} = T {\rm d}S - p {\rm d}V + \sum_{particles} \mu_i^{rel} {\rm d}N_i,
\end{eqnarray}
where
$E^{rel}$ is the total relativistic internal energy,
$S$ is the total entropy,
$V$ is the volume,
and
$N_i$ is the number of $i$-th particle contained in the system, respectively.
One may define a total baryon number in the system, $N_b \equiv \sum_{ion} A_i N_i$, as a constant value.
Then, specific densities of the relativistic internal energy, the entropy, the volume, and the number fractions are defined as
\begin{eqnarray}
	e^{rel}	&\equiv&	E^{rel}/ (m_u N_b) \\
	s		&\equiv&	S	/ (m_u N_b) \\
	1/\rho_b	&\equiv&	V	/ (m_u N_b) \\
	Y_i		&\equiv&	N_i	/ N_b.
\end{eqnarray}
Using these specific densities, the first law of thermodynamics in the specific density form
\begin{eqnarray}
	{\rm d}e^{rel} = T {\rm d}s - p {\rm d}\Bigl( \frac{1}{\rho_b} \Bigl) + \frac{1}{m_u} \sum_{particles} \mu_i^{rel} {\rm d}Y_i,
\end{eqnarray}
is obtained.


\bsp	
\label{lastpage}
\end{document}